\documentclass[reprint, superscriptaddress,amsmath,amssymb,aps,prl]{revtex4-1}

\usepackage{graphicx}
\usepackage{xcolor}
\usepackage{physics}
\usepackage{upgreek}
\usepackage{array}

\usepackage{multirow}
\newcommand{\boldgreek}[1]{{\mbox{\boldmath$ {#1} $}}}

\begin{document}
\title{Weak dissipation for high fidelity qubit state preparation and measurement}
\author{Anthony Ransford}
\email{ransfordanthony@gmail.com}
\affiliation{University of California Los Angeles, Los Angeles, CA, USA}
\author{Conrad Roman}
\author{Thomas Dellaert}
\author{Patrick McMillin}
\author{Wesley C. Campbell}
\affiliation{University of California Los Angeles, Los Angeles, CA, USA}
\affiliation{Challenge Institute for Quantum Computation, University of California Los Angeles, Los Angeles, CA, USA}

\date{\today}

\begin{abstract}

Highly state-selective, weakly dissipative population transfer is used to irreversibly move the population of one ground state qubit level of an atomic ion to an effectively stable excited manifold with high fidelity. Subsequent laser interrogation accurately distinguishes these electronic manifolds, and we demonstrate a total qubit state preparation and measurement (SPAM) inaccuracy $\epsilon_\mathrm{SPAM} < 1.7 \times 10^{-4}$ ($-38 \mbox{ dB}$), limited by imperfect population transfer between qubit eigenstates. We show experimentally that full transfer would yield an inaccuracy less than $8.0 \times 10^{-5}$ ($-41 \mbox{ dB}$).
The high precision of this method revealed a rare ($\approx 10^{-4}$) magnetic dipole decay induced error that we demonstrate can be corrected by driving an additional transition. Since this technique allows fluorescence collection for effectively unlimited periods, high fidelity qubit SPAM is achievable even with limited optical access and low quantum efficiency.

\end{abstract}

\maketitle

Recent progress in quantum device fidelity has focused primarily on improving unitary operations, \textit{i.e.}\ single and multi qubit gates, with some small systems achieving gate infidelities below thresholds necessary for fault tolerant encodings \cite{Knill1996ThresholdAF, Kitaev1997ErrorCorrection, Aharonov2008FaultTolerantQC}.
Despite these improvements, current systems lack the capacity to encode a computationally useful number of fault tolerant logical qubits. As such, current devices fall in the noisy intermediate scale quantum (NISQ) category \cite{Preskill2018NISQ}, where operations are performed without fault tolerance and the measurement fidelity of an $N$ qubit register will typically decrease exponentially with size as $(\mathcal{F}_\mathrm{SPAM})^N$.

Using strong dissipation ($\gamma/2 \pi = \mathcal{O}(\mbox{MHz})$) to ``shelve'' an electron to a metastable state \cite{dehmelt:1975}, the state preparation and measurement of a single qubit has recently been demonstrated with an infidelity $\epsilon_\mathrm{SPAM} < 3.2 \times 10^{-4}$ ($-35 \mbox{ dB}$) \cite{Christensen2020High}. The ultimate fidelity of this technique is limited by off-resonant coupling to strong electric dipole (E1) transition error channels during the shelving process and the finite lifetimes of the metastable states.  As an alternative to using strong transitions for population transfer, weak dissipative channels can also be used as a pathway to metastable states with a high degree of certainty \cite{roman_coherent_2020, Hayes2020LeakageRepump}. This can afford both highly selective transfer and high quality readout of the final qubit state \cite{roman_coherent_2020}.

In this Letter, we demonstrate and characterize the use of a weak dissipative channel in ${}^{171}\mathrm{Yb}^+$ hyperfine qubits to perform qubit state preparation and measurement with inaccuracy approaching $10^{-4}$. An electric quadrupole (E2) transition is driven by a laser to irreversibly transfer population from one qubit state to the effectively stable $^2\mathrm{F}^o_{7/2}$ state in $\mathcal{O}(100\mbox{ ms})$ \cite{roman_coherent_2020, Edmunds2021Scalable, Duan2021}. The long lifetime of this state ($\approx \! 2$ years \cite{Lange2021lifetime}) allows for laser-induced fluorescence to be collected for essentially unlimited duration without metastable decay, and we are able to distinguish the ground state from the metastable manifold with an inaccuracy $< \!-57\mbox{ dB}$ without high efficiency imaging.  The increased precision allotted by this technique revealed a qubit mixing error caused by a rare magnetic dipole (M1) decay ($A_\mathrm{M1} = 2\pi \times 4.1^{+2.3}_{-1.5} $ mHz) during population transfer, which we demonstrate can be corrected by introducing another laser beam.
We achieve a ground state hyperfine qubit SPAM inaccuracy $\epsilon_\mathrm{SPAM} = \!-39(1)\mbox{ dB}$, limited by the fidelity of unitary population transfer required to prepare one of the qubit states.
While narrowband optical pumping requires a longer duration (here, $\mathcal{O}(100\mbox{ ms})$) than a typical gate $\mathcal{O}(100\mbox{ }\mu \mbox{s})$, it is on par with the total algorithm times in current quantum systems \cite{Pino2021QCCD}, and is appropriate for initialization and readout of NISQ devices for which faulty SPAM will require the algorithm to be rerun.

\begin{figure}[t]
\begin{centering}
\includegraphics[width=0.9\linewidth]{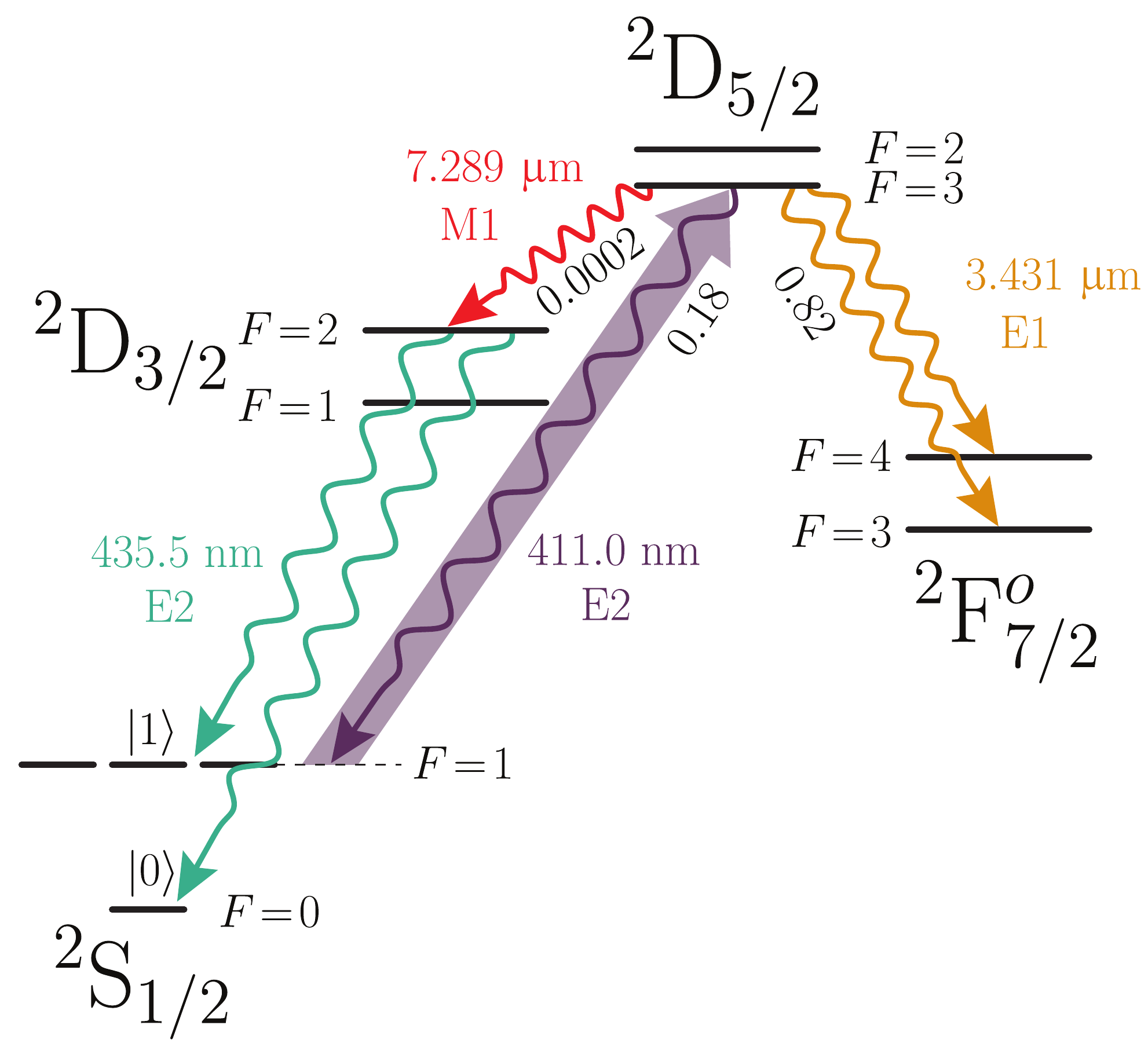}
\caption{\label{Histograms} Electron shelving of the $\ket{1}$ state via weak dissipation for high fidelity readout.  The M1 fine structure decay shown can induce a SPAM error due to its potential eventual decay to $\ket{0}$.  Each transition is labeled with its leading order identification as an electric dipole (E1), electric quadrupole (E2), or magnetic dipole (M1) process.}
\label{fig:figure1}
\end{centering}
\end{figure}

The weak dissipation scheme we present here requires a transition that is both narrow (for high state selectivity) and leaky (for robust, irreversible transfer).  The E2 transition in Yb$^+$ at 411 nm connecting the ground $^2\mathrm{S}_{1/2}$ state to $^2\mathrm{D}_{5/2}$ has a $\gamma = 2 \pi \times 22 \mbox{ Hz}$ linewidth, decays primarily to the ${}^2\mathrm{F}_{7/2}^o$ state, and has been investigated as a potential frequency standard \cite{Taylor1997ShelvingLine, Roberts1999DStateSplitting} and as a probe for physics beyond the standard model \cite{Counts2020SDNonlinearity}.
The extremely long lifetime of the $^2\mathrm{F}^o_{7/2}$ state and its optical isolation from the ${}^{171}\mathrm{Yb}^+$ cooling cycle makes it an ideal location to store qubit population during laser-induced fluorescence for state detection, and we use the 411 nm transition as the weakly open channel to this state, as shown in Figure \ref{fig:figure1}.

Selection rules allow for a quasi-cycling E2 transition on ${}^2\mathrm{S}_{1/2} (F\! =\! 1) \leftrightarrow {}^2\mathrm{D}_{5/2} (F\! =\! 3)$ to state selectively optically pump one $^{171}\mathrm{Yb}^+$ hyperfine level to $^2\mathrm{F}^o_{7/2}$ with E2-decay-induced mixing. The large hyperfine splitting of the intermediate ${}^2\mathrm{D}_{5/2}$ state ($191 \mbox{ MHz}$ \cite{Roberts1999DStateSplitting}) relative to $\gamma/2\pi$ minimizes the likelihood of off-resonant scattering, and the stability of the metastable state makes state misidentification due to spontaneous emission of the now hidden population a nonissue.
Following narrowband population transfer, subsequent laser interrogation reveals any population remaining in the $^2\mathrm{S}_{1/2}$ state with a high degree of certainty.

The ground state qubit is defined on the zero-field hyperfine clock states in $^{171} \mathrm{Yb}^+$, $\ket{^2\mathrm{S}_{1/2};F=0, M_F=0} \equiv \ket{0}$ and $\ket{^2\mathrm{S}_{1/2};F=1, M_F=0} \equiv \ket{1}$, with operating frequency $\omega_q = 2\pi \times 12.64 $ GHz. The ion is trapped in an oblate Paul trap  driven at $\Omega_\mathrm{RF}  =2\pi \times 49\mbox{ MHz}$ with typical secular frequencies of $(\omega_x, \omega_y, \omega_z) \approx 2\pi \times (540, 550, 1170)\mbox{ kHz} $ \cite{Yoshimura2015creation}. Light emitted by the ion in the $+z$ direction is focused by an off-the-shelf objective through an iris directly onto a PMT yielding a modest overall photon detection efficiency of $\approx$ 0.16\%. A magnetic field of $4.4 \mbox{ G}$ is applied in the $+z$ direction to lift the degeneracy of magnetically sensitive states. All lasers propagate in the $xy$ plane, perpendicular to the applied magnetic field. Qubit rotations are performed by microwave radiation delivered by an \textit{ex vacuo} standard gain horn antenna.

To evaluate the effectiveness of weak dissipative transfer of the $\ket{1}$ state to $^2\mathrm{F}^o_{7/2}$ for qubit measurement, we determine the fidelity of state preparation and electron shelving measurement of a single qubit by repeating SPAM attempts for each qubit basis state. Each experiment begins by Doppler cooling a single ion, after which laser light resonant with the $^2$S$_{1/2}(F=1) \leftrightarrow$ $^2\mathrm{P}^o_{1/2}(F=1)$ transition is applied to prepare the $\ket{0}$ state with high fidelity \cite{Olmschenk2007Manipulation}. After preparation of $\ket{0}$, resonant microwaves can used to transfer population to $\ket{1}$ for $\ket{1}$ state SPAM. This transition frequency is calibrated periodically by performing Ramsey spectroscopy, and we find that the relative qubit-oscillator frequency drift is typically less than $20 \mbox{ Hz}$ over 24 hours. The microwave interrogation times are calibrated every 2000 experiments by fitting Rabi flops. 

Once state preparation is complete, $2 \mbox{ mW}$ of $411 \mbox{ nm}$ laser light is directed onto the ion at a spot size of $80 \mbox{ }\upmu\mbox{m}$ in order to transfer population in the $\ket{1}$ state to the $^2\mathrm{F}^o_{7/2}$ states through the $^2\mathrm{D}_{5/2}(F=3)$ manifold. 
Light at $935 \mbox{ nm}$ is also applied to repump any population in the $^2\mathrm{D}_{3/2}$ state. The $411 \mbox{ nm}$ and $935 \mbox{ nm}$ light is applied for $200 \mbox{ ms}$, long enough to ensure a population transfer infidelity of $<-50 \mbox{ dB}$ out of $^2\mathrm{S}_{1/2}(F=1)$. The polarization of the $411 \mbox{ nm}$ light is chosen to maximize the transition strengths of the $\abs{\Delta M_F} = 2$ transitions ($\mathbf{k} \perp \boldgreek{\hat{\varepsilon}} \perp \mathbf{B}$). We transfer using the $\Delta M_F = -2$ transition due to its larger detuning from $^2\mathrm{D}_{5/2}(F=2)$. Since the $g$-factors of the $^2\mathrm{S}_{1/2}(F=1)$ and $^2\mathrm{D}_{5/2}(F=3)$ manifolds are neary equal ($\Delta g_F \approx 10^{-3}$), all magnetic sublevels in $^2\mathrm{S}_{1/2} (F=1)$ are coupled to ${}^2\mathrm{D}_{5/2}(F=3)$ with a single laser frequency.

After electron shelving, we detect any remaining ${}^2\mathrm{S}_{1/2}$ population by laser-induced fluorescence for $17 \mbox{ ms}$. Any fluorescence photons collected from the ion are counted and timetagged with 10 ns precision by a custom FPGA-based pulse sequencer \cite{Pruttivarasin2015compact}. Population that has been shelved to the $^2\mathrm{F}^o_{7/2}$ manifold does not produce laser-induced fluorescence. 

Following state detection, population in $^2\mathrm{F}^o_{7/2}$ is ``deshelved'' by driving the E2 transition to ${}^1[3/2]_{3/2}^o$ at $760 \mbox{ nm}$ for $35 \mbox{ ms}$. Due to the hyperfine structure in the $^2\mathrm{F}^o_{7/2}$ and the excited $^1[3/2]^o_{3/2}$ states, two $760 \mbox{ nm}$ tones separated by $5.257 \mbox{ GHz}$ are applied for quick depopulation \cite{Mulholland_2019}. 
To ensure efficient return to the ground state during deshelving, lasers at $976 \mbox{ nm}$ and $935 \mbox{ nm}$ are used to depopulate the $^2\mathrm{D}_{5/2}$ and $^2\mathrm{D}_{3/2}$ states via ${}^1[3/2]_{3/2}^o$ and ${}^3[3/2]_{1/2}^o$, respectively. Population is returned to the ground state manifold with a $1/e$ time of $\tau \approx 350 \mbox{ }\upmu \mbox{s}$. During deshelving, the lasers used for Doppler cooling are also applied. 

Data is collected by interleaving sets of 50 SPAM attempts for each qubit state. A single experiment consists of preparing the $\ket{0}$ ($\ket{1}$) qubit state, shelving population in the ${}^2\mathrm{S}_{1/2}(F=1)$ level, querying fluorescence from the $^2\mathrm{S}_{1/2}$ manifold, and then deshelving the population in the $^2\mathrm{F}^o_{7/2}$ manifold. 

Throughout the experimental run we monitor the number of photons collected during Doppler cooling and use a threshold to restart experiments where an ion was not properly cooled \emph{prior to} a state preparation and measurement attempt.  Experiments where only the Doppler cooling counts \emph{following} the SPAM attempt fall below the threshold are identified as ion storage errors.  We reserve the use of the term ``infidelity'' to include storage errors, as those events are clear failures to perform what was intended and can only be identified as failures after the fact.  However, since these storage errors are flagged by low Doppler cooling counts after the SPAM attempt, they do not result in a misidentification of the qubit state, and we use the term ``inaccuracy'' to refer to SPAM errors other than ion storage errors.

To experimentally measure the SPAM inaccuracy, we adhered to a blinded data analysis to avoid introducing bias when choosing the thresholds for Doppler cooling and state discrimination. Prior to the measurement of the final data set, $\sim 10^4$ experiments per qubit state were performed, and appropriate thresholds for state discrimination and Doppler cooling were set and fixed based on those calibration data. The state discrimination threshold is chosen such that a successful preparation of the $\ket{1}$ state would produce an error with probability $\leq 10^{-7}$ based on the $\ket{1}$ state distribution mean of the calibration data. The Doppler cooling threshold is similarly chosen such that a properly cooled ion would fail the thresholding with probability $\leq 10^{-6}$ based on Doppler cooling count rate. Once these thresholds were fixed, the final data set was then unblinded and analyzed, resulting in $5\times 10^4$ data points per state after removing ion storage errors.

\begin{figure}[t]
\begin{centering}
\includegraphics[width=\linewidth]{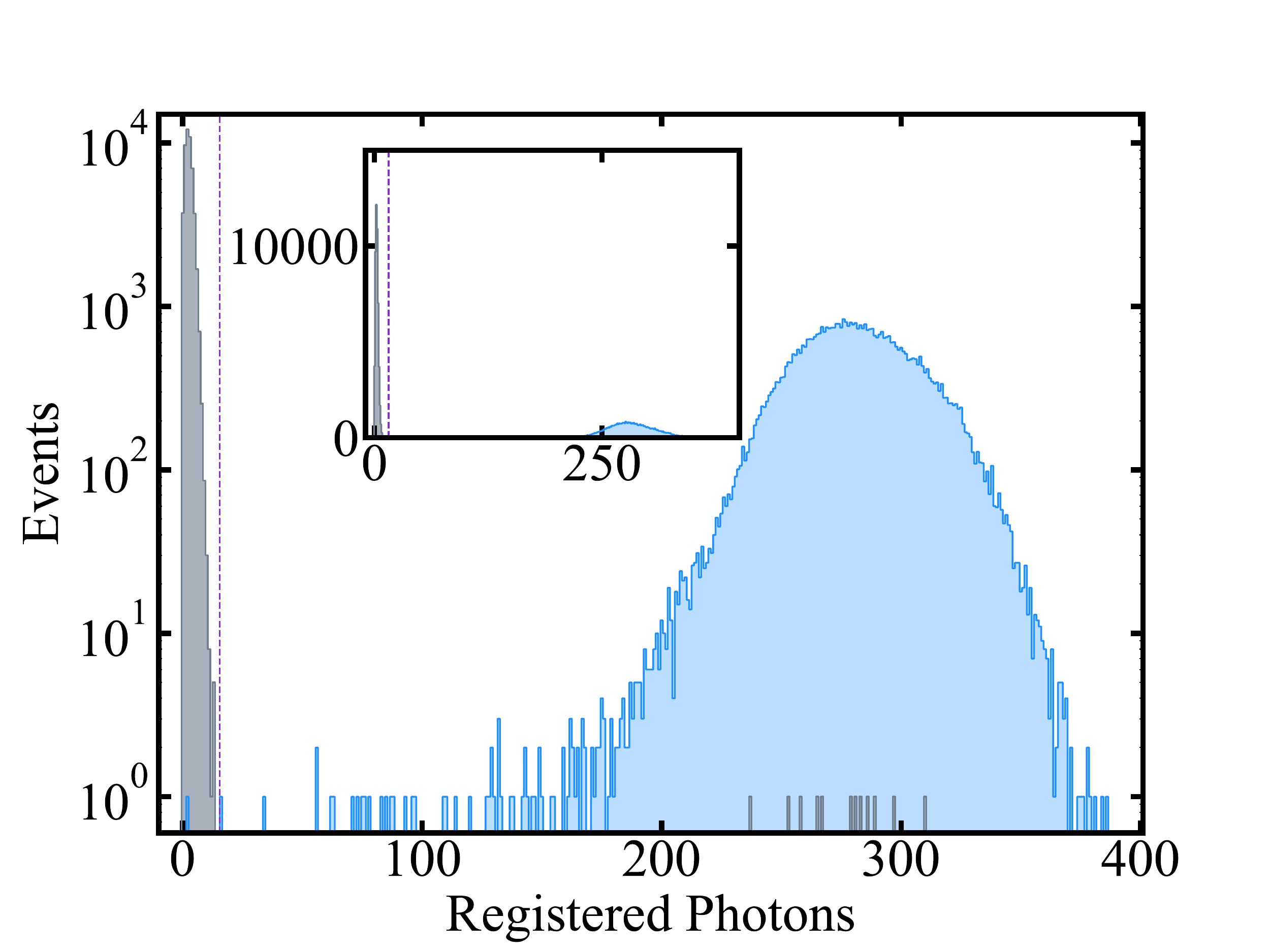}
\caption{Fluorescence detection count histograms for attempted SPAM of the $\ket{1}$ (blue) and $\ket{0}$ (gray) ground state hyperfine qubit states in ${}^{171}\mathrm{Yb}^+$.  The inset shows the same data on a linear scale. The predetermined state detection threshold is shown as a dashed purple line, and gives an average SPAM inaccuracy of $-39(1)\mbox{ dB}$.}
\label{fig:histogram}
\end{centering}
\end{figure}

\begin{figure}[t]
    \centering
    \includegraphics[width=\linewidth]{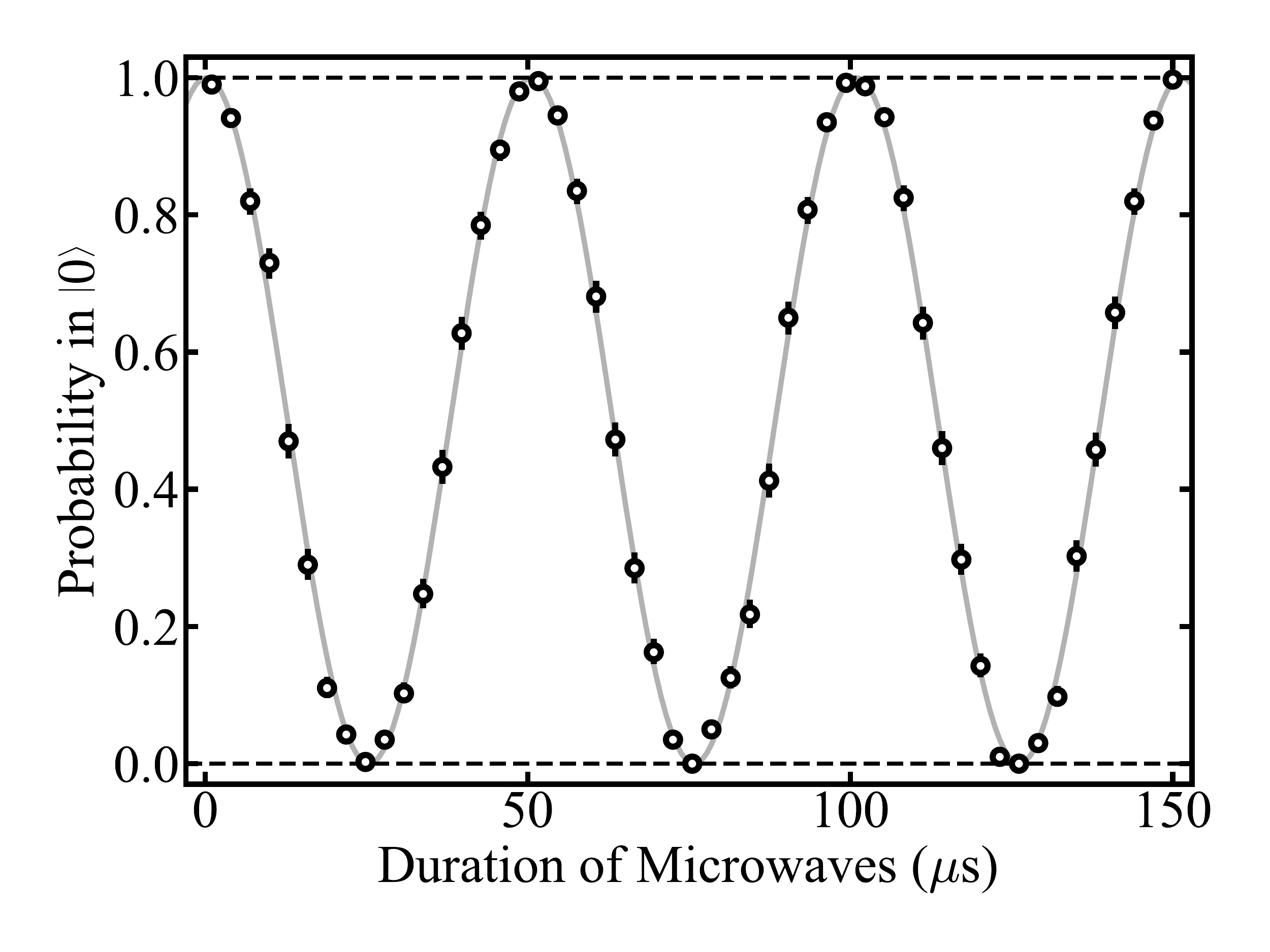}
    \caption{High contrast readout of microwave rotations of a ${}^{171}\mathrm{Yb}^+$ ground state hyperfine qubit measured via electron shelving with $\approx 400$ measurements per point.}
    \label{fig:rabi_flop}
\end{figure}

Following this procedure, we observe a total state preparation and measurement inaccuracy  $\epsilon_\mathrm{SPAM} =-39(1)\mbox{ dB}$ ($1.3^{+0.4}_{-0.3} \times10^{-4}$), where the uncertainty is a one sigma Wilson score interval.  (The SPAM infidelity, which counts ion storage errors as failures, is $1-\mathcal{F}_\mathrm{SPAM} = -37(1) \mbox{ dB}$.)  These results are consistent with the prediction of the error budget sown in Table \ref{table:errorbudget}. The photon count histograms are shown in Figure \ref{fig:histogram} and show clear separation of the two distributions, illustrating that the SPAM accuracy is not limited by the ability to distinguish the ${}^2\mathrm{S}_{1/2}$ manifold from the ${}^2\mathrm{F}_{7/2}^o$ manifold.  Figure \ref{fig:rabi_flop} shows the high contrast of qubit Rabi flopping that can be detected using this method.

The sources of the SPAM inaccuracy are asymmetric between the two qubit states. The SPAM inaccuracy of the $\ket{0}$ state is found be $-47(4)\mbox{ dB}$, roughly $10\times$ lower than that of the $\ket{1}$ state, $-36(1)\mbox{ dB}$.
The asymmetry comes from two sources: imperfect microwave transfer on $\ket{0} \rightarrow \ket{1}$ causing preparation of the $\ket{1}$ state to fail, and spontaneous M1 decay during shelving from $^2\mathrm{D}_{5/2} (F = 3)$ to $^2\mathrm{D}_{3/2} (F = 2)$ that subsequently decays to $\ket{0}$, causing shelving to fail.

We quantified the quality of our microwave rotations by performing single qubit gate randomized benchmarking \cite{Knill2008RandomizedBenchmarking}, finding that the infidelity of our randomized single computational gates is $\epsilon_{\pi}  = -41.3(6) \mbox{ dB}$ \cite{SupplementalMaterials}.  In the final data set, by subtracting the amount of shelving error we expect to be contributed from the M1 decay and the finite shelving time, we find that the remaining error is $-39(2) \mbox{ dB}$, which we attribute to imperfections in this transfer. 

\begin{figure}[t]
\centering
\includegraphics[width=1\linewidth]{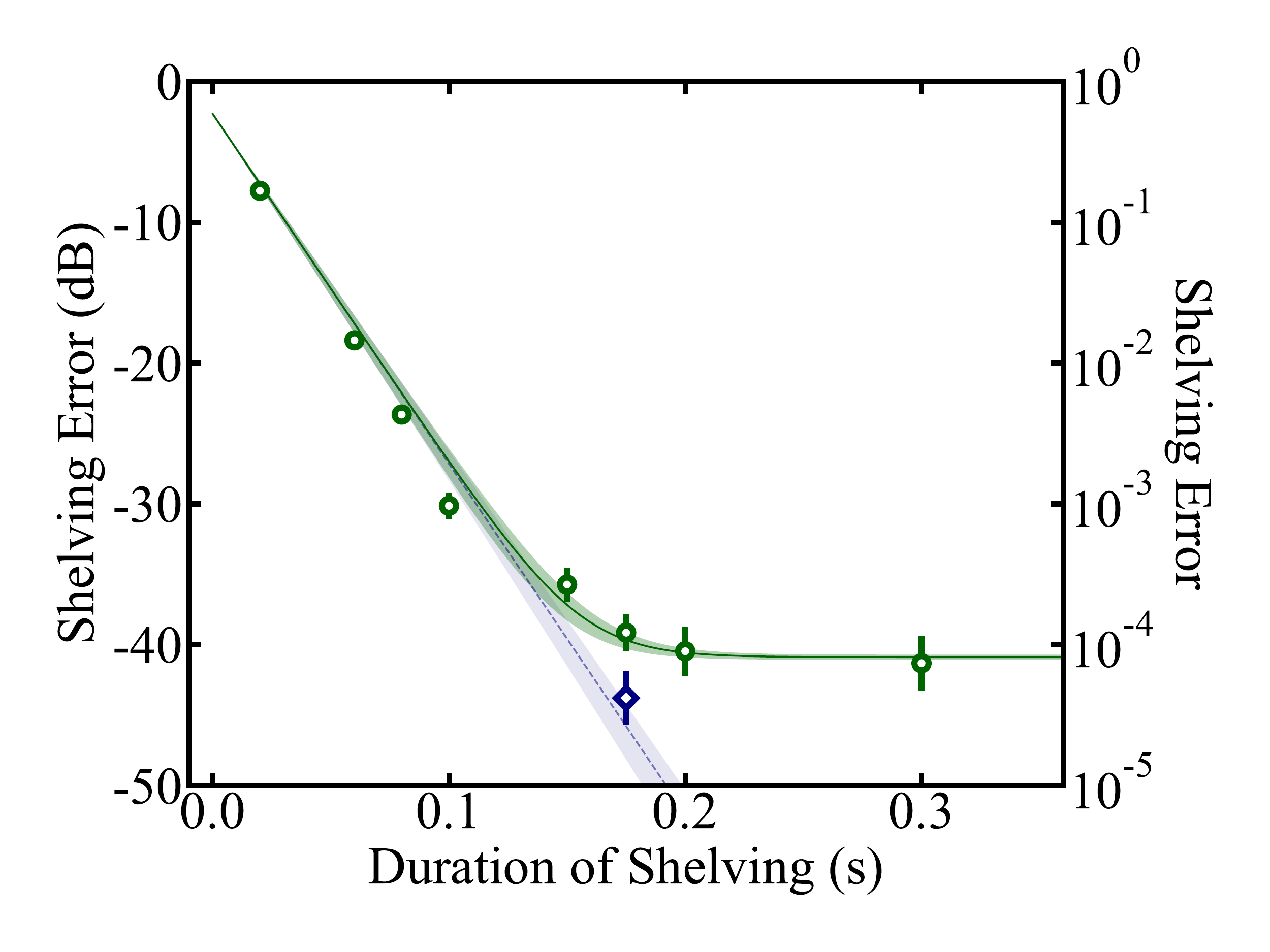}
\caption{Error in electron shelving of an ion prepared initially in the ${}^2\mathrm{S}_{1/2}(F=1)$ manifold as a function of $411\mbox{ nm}$ illumination times.  The green trace and circles show the error with only the $935 \mbox{ nm}$ light repumping population from the ${}^2\mathrm{D}_{3/2}$ states, while the blue trace and diamond shows the error with the $935 \mbox{ nm}$ light replaced by the $861 \mbox{ nm}$ repump scheme. Dashed lines are the theoretical prediction of Eq.~(\ref{eq:shelvingcurve}), with bands indicating the uncertainty in the model due to uncertainty in the ${}^2\mathrm{D}_{5/2}$ lifetime and E1-E2 branching ratio.  The prediction with the $861 \mbox{ nm}$ repump (blue) is just the second term in Eq.~(\ref{eq:shelvingcurve}).} 
\label{fig:shelving_function_time}
\end{figure}

With higher fidelity state preparation of the $\ket{1}$ state, the state detection error from the fine structure M1 transition becomes the dominant source of SPAM inaccuracy for transfer times longer than $150 \mbox{ ms}$. The ${}^2\mathrm{S}_{1/2}(F=1)$ manifold can be prepared with higher probability than just its $\ket{M_F=0}$ quantum state by performing a series of three $\pi$-rotations from $\ket{0}$ to the three ${}^2\mathrm{S}_{1/2}(F=1)$ magnetic sublevels. When illuminated by $411 \mbox{ nm}$ laser light, the entire ${}^2\mathrm{S}_{1/2}(F=1)$ manifold will be shelved since the  ${}^2\mathrm{D}_{5/2}(F=3)$ manifold has an equal Land\'e $g$-factor. Figure \ref{fig:shelving_function_time} shows the measured probability of finding an unshelved ion, if it is first prepared in the ${}^2\mathrm{S}_{1/2}(F=1)$ manifold, as a function of the shelving illumination time. The green circles show the measured error with the $935 \mbox{ nm}$ repump light on during shelving (which only partially protects against decay to $\ket{0}$ during the shelving of a $\ket{1}$ ion). The shelving error with this scheme is given approximately by
\begin{equation}
    \epsilon_\mathrm{s}(t) =  \frac{\tau_\mathrm{D} A_\mathrm{M1}}{3\zeta} + \left(1 - \frac{\zeta}{2} \right)\exp \left(-\frac{t \zeta}{2 \tau_\mathrm{D}} \right),\label{eq:shelvingcurve}
\end{equation}
which assumes $\tau_\mathrm{D} A_\mathrm{M1} \ll 1$ and that all of the ${}^2\mathrm{D}_J$ population has decayed before fluorescence querying.  Here, $\zeta = 0.824(4)$ is the branching ratio to the $^2\mathrm{F}^o_{7/2}$ manifold \cite{Tan2021Precision}, and $\tau_\mathrm{D} = 7.2(3)\mbox{ ms}$ is the lifetime of the $^2\mathrm{D}_{5/2}$ state \cite{Taylor1997ShelvingLine}. Using a calculated value for $A_\mathrm{M1} = 2 \pi \times 4.5\mbox{ mHz}$ \cite{SupplementalMaterials} yields a predicted error contribution of $\epsilon_\mathrm{M1} = \epsilon_\mathrm{s} (t\rightarrow \infty) =  -40.8(2)\mbox{ dB}$, where the uncertainty comes from the uncertainty in $\tau_\mathrm{D}$. The presence of this underappreciated decay channel highlights the importance of measuring and including the shelving error when reporting state detection errors, as it is not possible to achieve qubit state detection error lower than $\approx -40 \mbox{ dB}$ with this scheme without somehow addressing this source of infidelity.

\begin{table}[]
\begin{tabular}{|l|>{\centering\arraybackslash}m{0.2\columnwidth}|>{\centering\arraybackslash}m{0.2\columnwidth}|}
\hline
\multicolumn{1}{|c|}{\multirow{2}{*}{Error Source}} & \multicolumn{2}{c|}{Predicted Error ($\times 10^{-4}$)} \\
\multicolumn{1}{|c|}{}                    & $\ket{1}$ State       & $\ket{0}$ State            \\ \hline\hline
$\ket{0}$ state preparation               & \multicolumn{2}{c|}{$<0.02$}                       \\ \cline{2-3} 
Unflagged storage error                   & ---                   & $0.1^{+0.2}_{-0.06}$       \\ \cline{2-3} 
$\ket{0} \rightarrow \ket{1}$ transfer    & $0.74(10)$            & ---                        \\ \cline{2-3} 
Finite shelving time                      & $0.06(3)$             & ---                        \\ \cline{2-3} 
M1 decay                                  & $0.82(3)$             & ---                        \\ \hline 
\textbf{Predicted average inaccuracy}     & \multicolumn{2}{c|}{$\mathbf{0.9^{+0.2}_{-0.1}}$}  \\ \hline\hline
Flagged storage error                     & ---                   & $2.9^{+0.6}_{-0.5}$        \\ \hline 
\textbf{Predicted average infidelity}     & \multicolumn{2}{c|}{$\mathbf{2.4^{+0.4}_{-0.3}}$}  \\ \hline
\end{tabular}
\caption{Error budget for SPAM measurement determined from theoretical estimates and auxiliary measurements.
$\ket{0}$ state preparation errors are common to both the $\ket{0}$ state and the $\ket{1}$ state, while the other sources of error apply only to either the $\ket{1}$ or the $\ket{0}$ state.\label{table:errorbudget}}  
\end{table}

The magnetic transition dipole moment between the ${}^2\mathrm{D}_J$ levels can be measured by counting shelving errors caused by this decay using the same procedure, and the infidelity can be traced entirely to the M1 decay pathway at a shelving time of $300 \mbox{ ms}$.  This measured error yields an M1 transition rate of $A_\mathrm{M1} = 2\pi \times 4.1^{+2.3}_{-1.5}$ mHz, consistent with the theoretical prediction of  $2\pi \times 4.5 $ mHz \cite{SupplementalMaterials}. This corresponds to a branching ratio of $1.8^{+1.0}_{-0.7} \times 10^{-4}$ for the ${}^2\mathrm{D}_{5/2} \rightarrow {}^2\mathrm{D}_{3/2}$ decay channel.

This error channel does not set a fundamental limit to this method, and can be further suppressed with the addition of light that repumps this population via a higher angular momentum state. Laser light at $861 \mbox{ nm}$ or $1.35 \mbox{ }\upmu \mbox{m}$ can be used to state-selectively depopulate the $^2\mathrm{D}_{3/2} (F = 2)$ level through $^1[3/2]^o_{3/2} (F = 2)$ or $^2\mathrm{P}^o_{3/2} (F = 2)$ respectively, both of which decay via E1 transitions quickly and predominantly to the $^2\mathrm{S}_{1/2}(F=1)$ manifold. 
The blue diamond in Fig.~\ref{fig:shelving_function_time} shows the measured effect on manifold readout of adding an 861 nm laser (and removing the $935 \mbox{ nm}$ light) to mitigate the M1 decay error for a shelving time of $175 \mbox{ ms}$. We find that the measured shelving error is reduced by about $4\mbox{ dB}$ by this repump scheme. A binomial test applied to these two data points confirms this as a statistically significant suppression of the error mode ($p=0.0038$). 

We also measure how well we can distinguish ions in the $^2\mathrm{S}_{1/2}$ manifold from ions in the $^2\mathrm{F}^o_{7/2}$ manifold. This is by far the most straightforward part of this technique and should not be confused with qubit SPAM fidelity, which must include all of the error sources described above.  We perform this measurement by laser cooling and counting photons produced by two ions that are either both prepared in the $^2\mathrm{S}_{1/2}$ manifold or with precisely one excited to the ${}^2\mathrm{F}^o_{7/2}$ manifold. Photons are counted in $10 \mbox{ ms}$ bins, with every even bin taken as a detection bin, and the complementary odd bins taken as a check for a properly cooled ion crystal before and after a detection bin to check for and eliminate storage errors from the data set. We find that we can distinguish an ion in the metastable manifold from an ion in the ground state with an inaccuracy $ \epsilon_\mathrm{S/F} < -57 \mbox{ dB}$.

One of the attractive features of the $411 \mbox{ nm}$ electron shelving qubit readout scheme in ${}^{171}\mathrm{Yb}^+$ \cite{roman_coherent_2020} is the practically unlimited number of fluorescence photons that bright ions can emit.  This capability can boost statistical rejection of fluorescence cross-talk due to the overlap of an imaging system's point spread functions from neighboring ions in a Coulomb crystal.  However, care must be taken to ensure that the single $\lambda = 3.4 \mbox{ }\upmu\mbox{m}$ photon that must be spontaneously emitted by each ion being shelved does not de-shelve neighboring ions, as the resonant absorption cross section ($\mathcal{O}(\lambda^2)$) spans a length scale that may be similar to the inter-ion spacing.  In this case, it should be possible to utilize the AC Stark shift from continuous $411 \mbox{ nm}$ illumination to make each ion's $3.4 \mbox{ }\upmu\mbox{m}$ resonance frequency unique to avoid superradiant and reabsorption effects
 
Improving the speed at which qubit population can be state selectively transferred to the $^2\mathrm{F}^o_{7/2}$ and reducing the required laser intensity required to return $^2\mathrm{F}^o_{7/2}$ state population to the laser cooling cycle are two important areas in which this scheme can be improved. To improve shelving speed, qubit population can be coherently transferred to the metastable state instead of relying on the relatively slow E1 decay at 3.4 $\upmu$m. This can be done either directly at 467 nm \cite{E3ClockPTB2021}, or using both $411 \mbox{ nm}$ and $3.4 \mbox{ }\upmu \mbox{m}$ light, as was recently demonstrated along with high speed, low intensity depopulation of the $^2\mathrm{F}^o_{7/2}$ state with $3.4 \mbox{ }\upmu \mbox{m}$ and $976 \mbox{ nm}$ light \cite{Duan2021}.

\begin{acknowledgments}
The authors acknowledge Justin Christensen, David Hucul, Nils Huntemann for helpful discussions. This work was partially supported by the U.S.~Army Research Office under Grant No. W911NF-15-1-0261 and the U.S. National Science Foundation under Award Nos.\ PHY-1912555 and OMA-2016245.
A.~R.\ and C.~R.\ contributed equally to this work.
\end{acknowledgments}

\bibliography{Fshelving}
\end{document}